\documentclass[12pt]{article}
\usepackage{amsmath}
\usepackage{graphics,graphicx,epsfig}
\usepackage{amssymb}
\usepackage{epstopdf}
\usepackage{sidecap}
\usepackage{appendix}
\usepackage{cite}
\usepackage[hang,flushmargin]{footmisc}

\begin{document}

\begin{titlepage}
\vfill
\begin{flushright}
ACFI-T18-09
\end{flushright}

\vfill
\begin{center}
\baselineskip=16pt
{\Large\bf Black Hole Enthalpy and Scalar Fields}
\vskip 0.15in

{\bf  David Kastor${}^{a}$, Sourya Ray${}^{b}$, Jennie Traschen${}^{a}$} 

\vskip 0.4cm
${}^a$ Amherst Center for Fundamental Interactions, Department of Physics\\ University of Massachusetts, Amherst, MA 01003\\

\vskip 0.1in ${}^b$ Instituto de Ciencias F\'{\i}sicas y Matem\'{a}ticas, Universidad Austral de
Chile, Valdivia, Chile\\
\vskip 0.1 in Email: \texttt{kastor@physics.umass.edu, ray@uach.cl, traschen@physics.umass.edu}
\vspace{6pt}
\end{center}
\vskip 0.2in
\par
\begin{center}
{\bf Abstract}
 \end{center}
\begin{quote}
The mass of an AdS black hole represents its enthalpy, which in addition to internal energy, generally includes the energy required to assemble the system in its environment.  In this paper, we consider black holes immersed in a more complex environment, generated by a scalar field with an exponential potential.  In the analogue of the AdS vacuum, which we call dilaton-AdS, the scalar field has non-trivial behavior, breaking the hyperscaling symmetry of AdS and modifying the asymptotic form of the spacetime.  We find that the scalar field falloff at infinity makes novel contributions to the ADM mass and spatial tensions of dilaton-AdS black holes.  We derive a first law and Smarr formula for planar dilaton-AdS black holes.  We study the analogue of thermodynamic volume in this system and show that the mass of a black hole again represents its enthalpy.
\vfill
\vskip 2.mm
\end{quote}
\hfill
\end{titlepage}


\section{Introduction}

It is interesting to understand how a black hole is affected by changes to its external environment.  A simple example of an environmental variable, which we can consider tuning at least in theory, is a non-zero cosmological constant $\Lambda$.  For AdS black holes ({\it i.e.} with $\Lambda<0$) it was shown \cite{Kastor:2009wy} following earlier observations \cite{Caldarelli:1999xj,Wang:2006eb,Sekiwa:2006qj,Wang:2006bn,Larranaga Rubio:2007jz,Urano:2009xn} that when $\Lambda$ is varied, the variations in the black hole mass $M$ and horizon area $A$ are related via the extended first law 
\begin{equation}\label{adsfirst}
\delta M = {\kappa \delta A\over 8\pi} +{\Theta \delta\Lambda\over 8\pi}
\end{equation}
where $\kappa$ is the horizon surface gravity and $\Theta$ is a new thermodynamic potential with dimension of spatial volume.  For spherical black holes, the cosmological constant also contributes to the Smarr formula  \cite{Kastor:2009wy}
\begin{equation}\label{adssmarr}
(D-3)M = (D-2) {\kappa A\over 8\pi} - 2{\Theta \Lambda\over 8\pi}
\end{equation}
where the numerical factors correspond to the scaling of the quantities $M$, $A$ and $\Lambda$, which have dimensions of length to the power $D-3$, $D-2$, and $-2$ respectively.  These formulas show the impact of $\Lambda$ on the relation between $M$, which is defined at spatial infinity, and $\kappa$ and $A$, which are properties of the black hole horizon.

The physical significance of $\Theta$ is clarified by thinking of the cosmological constant as a pressure $P\equiv -\Lambda/8\pi$ together with a conjugate `thermodynamic  volume' $V=-\Theta$.  The extended first law can then be written as
\begin{equation}\label{firstlaw}
\delta M = T\delta S+V\delta P
\end{equation}
where, as usual, $T=\kappa/2\pi$ and $S=A/4$ are the black hole temperature and entropy.
In terms of ordinary thermodynamics, one sees that $M$ appears as the black hole enthalpy \cite{Kastor:2009wy}, which includes in addition to internal energy, a contribution of $PV$ necessary to create the black hole in its environment.

For spherical Schwarzschild-AdS black holes, the thermodynamic volume $V$ is equal to the Euclidean volume of a sphere of the horizon radius $r_h$.  Thermodynamic volumes were computed for a wide range of AdS black holes including charge and rotation \cite{Cvetic:2010jb} leading to a conjectured `reverse isoperimetric inequality', that bounds thermodynamic volume from below in terms of the black hole entropy.  A study of deSitter black holes \cite{Dolan:2013ft} led to a pair of independent extended first laws. In addition to a first law like (\ref{firstlaw}), a second one corresponding to the region between the black hole and deSitter horizons relates variations in the black hole and cosmological horizon areas to $\delta\Lambda$, without the presence of a $\delta M$ term.  

Considering an extended phase space for black holes that includes $\Lambda$ as a thermodynamic variable has led to a number of further developments, including discussion of phase transitions (beginning with \cite{Kubiznak:2012wp}; see \cite{Kubiznak:2016qmn} for a review and extensive references), and  analysis of the efficiency of `holographic heat engines' (begining with \cite{Johnson:2014yja}).    At the quantum mechanical level, it has been suggested that the thermodynamic volume may be related to the quantum complexity of the black hole \cite{Couch:2016exn}.

The richness of these results motivates our focus in this paper on a somewhat more complex environment for a black hole, namely one involving a scalar field.  More specifically, we consider black holes in Einstein gravity coupled to a scalar field with an exponential potential and described by the action 
\begin{equation}\label{action}
S = \int d^{n+1}x \sqrt{-g}\left( R-{1\over 2}(\nabla\phi)^2-2\Lambda e^{-a\phi}\right)
\end{equation}
in $D=n+1$ dimensions.  The theory includes a dimensionless coupling constant $a$, such that for $a=0$ the interaction reduces to the  cosmological constant case discussed above.  We focus here on $\Lambda<0$, because of its close relation with AdS/CFT.  However 
$\Lambda>0$ is also of physical interest, because the analogues of the planar black holes discussed here will then be cosmological models that asymptote at future infinity to power law inflationary cosmologies\footnote{See \cite{Kastor:2016bnm} for a discussion of the asymptotically deSitter cosmological case.}.  Also without loss of generality, we will consider only $a\ge 0$.

Before considering black hole solutions, we first need to identify appropriate vacuum solutions to (\ref{action}).  For our study,  we take these to be 
\begin{equation}\label{dilatonAdS}
ds^2= \left({l^2\over r^2}\right)^{1-\sigma}dr^2+{r^2\over l^2}\,\eta_{\alpha\beta}dx^\alpha dx^\beta,\qquad e^{a\phi} = \left({r^2\over l^2}\right)^{\sigma},\qquad \sigma={1\over 2}(n-1)a^2
\end{equation}
where $\eta_{\alpha\beta}$ with $\alpha,\beta = 0,1,\dots,n-1$ is the Minkowski metric, 
and the length scale $l$ appearing in the metric is related to the strength $\Lambda$ of the scalar field potential by
\begin{equation}\label{dlambda}
\Lambda = -{(n-1)(n-\sigma)\over 2l^2}
\end{equation}
For $a=0$ this reduces to AdS spacetime in Poincare coordinates.  We will refer to the solutions (\ref{dilatonAdS}) as dilaton-AdS  spacetimes, and the black hole solutions we introduce below will be asymptotically dilaton-AdS at spatial infinity.  Asymptotically dilaton-AdS spacetimes, also including an electromagnetic field coupled to the dilaton, have been extensively  studied in \cite{Charmousis:2010zz}\footnote{One can alternatively study black holes asymptotic at infinity to a static spherically symmetric vacuum solution as in reference \cite{Wiltshire:1990ah}.}. 
Note that the scalar field does not vanish at infinity in  the dilaton-AdS vacua (\ref{dilatonAdS}).  We will see below that this leads to a novel scalar field contribution to the ADM mass for asymptotically dilaton-AdS spacetimes.

The dilaton-AdS vacua (\ref{dilatonAdS}) have Poincare symmetry along the slices at constant radius.  Another key feature is the presence of a homothetic Killing vector \cite{Dong:2012se}
\begin{equation}\label{hkv}
\xi= r{\partial\over\partial r}-(1-\sigma)x^\alpha{\partial\over\partial x^\alpha}
\end{equation}
generating a scaling transformation $r\rightarrow \lambda r$ and $x^\alpha\rightarrow \lambda^{-(1-\sigma)} x^\alpha$
under which the metric is rescaled by a constant conformal factor, $ds^2\rightarrow \lambda^{-2\sigma}ds^2$.  For $a=0$ 
this is an exact isometry of AdS spacetime.  In the context of AdS/CFT this bulk spacetime symmetry underlies the overall scale invariance of the boundary CFT \cite{Maldacena:1997re}.  
This exact isometry also implies a corresponding trace-free condition 
for the ADM mass and spatial tensions of asymptotically planar AdS spacetimes \cite{El-Menoufi:2013tca}.
The failure of precise scaling isometry for $a\neq 0$ is known as hyperscaling violation.

We will explore how the extended phase space approach, including the overall strength $\Lambda$ of the scalar potential as a thermodynamic variable, works in the presence of such hyperscaling violation\footnote{Hyperscaling violation is often studied in conjunction with Lifshitz-type anisotropy between the scaling of time and space dimensions \cite{Kachru:2008yh}.  This requires more complicated matter Lagrangians.  Here we focus on the effects of hyperscaling violation and leave the issue of Lifshitz scaling for future work.  However, see \cite{Brenna:2015pqa,Ma:2017pap} for work on thermodynamic volume in this context.}.
We will derive a first law and a Smarr formula for planar dilaton-AdS black holes that are analogous to (\ref{adsfirst}) and (\ref{adssmarr}), but with one addition.  It was emphasized in \cite{Kastor:2009wy} that a Smarr formula can be derived via scaling from a first law that is extended to include variations in all relevant dimensionful quantities, such as the cosmological constant.  For planar black holes, in order to make the ADM charges finite, we must take the spatial directions parallel to the horizon to be compact.  Accordingly, we will make the identifications\footnote{More generally one could compactify on a non-orthogonal torus, which would introduce additional moduli.} $x^k\equiv x^k+ L_k$ for $k=1,\dots,n-1$.  The compactification lengths $L_k$ are then additional dimensionful parameters whose variations must be included in a first law for planar black holes
\begin{equation}\label{planarfirstlaw}
\delta M = {\kappa \delta A\over 8\pi} +\sum_{k=1}^{n-1} {\cal T}_k \delta L_k +{\Theta \delta\Lambda\over 8\pi}
\end{equation}
where ${\cal T}_k$ is the ADM tension in the $k$th spatial direction\footnote{See \cite{Kastor:2006ti} for a derivation of the tension contributions to the first law in the context of black branes with $\Lambda=0$.  This also includes discussion of a case in which the compactification torus cycles are allowed to be non-orthogonal, in which case there is a spatial tension tensor.}.  We will argue that the last term in (\ref{planarfirstlaw}) may again be regarded as a $V\delta P$ term, and consequently that $M$ should be regarded as the black hole enthalpy, as in the AdS case.  Since the compactification lengths have scaling dimension $(length)^1$, the associated Smarr relation is
\begin{equation}\label{planarsmarr}
(D-3)M = (D-2) {\kappa A\over 8\pi} +\sum_{k=1}^{n-1} {\cal T}_k L_k - 2{\Theta \Lambda\over 8\pi}
\end{equation}
This Smarr formula was derived for planar AdS black holes in \cite{El-Menoufi:2013pza}\footnote{The associated first law for planar black holes was not discussed in \cite{El-Menoufi:2013pza}, but follows from the $a=0$ limit of the results in the present paper.}.  We will find a similar relation here.  However, we will see that the derivation of the thermodynamic potential $\Theta$ arising from the scalar field potential in (\ref{action}) is somewhat more complicated than in the AdS case.

The paper proceeds as follows.  In section (\ref{admsection}) we compute the mass and spatial tensions of spacetimes that are asymptotic at large distance to the dilaton-AdS vacuum solutions (\ref{dilatonAdS}). 
In section (\ref{dadsbhsection}) we present a family of black hole solutions with regular, planar horizons that are asymptotic to the dilaton-AdS spacetimes (\ref{dilatonAdS}) at spatial infinity.  These black hole solutions are a subset of a larger class of `Kasner-dilaton-AdS' (KDAdS) solutions to (\ref{action}), which generically have singular horizons.  In section (\ref{smarrsection}) we use Komar integrals to derive a Smarr formula for dilaton-AdS black holes.  In section (\ref{firstlawsection}) we derive the first law for dilaton-AdS black holes.  A scaling argument can then also be used to reproduce the Smarr formula of the previous section.  In section (\ref{phenosection}) we check that the dilaton-AdS black hole solutions found in section (\ref{dadsbhsection}) indeed satisfy this first law and Smarr formula.  Section (\ref{conclusions}) contains some concluding remarks and directions for future work.

\section{Asymptotically dilaton-AdS spacetimes}\label{admsection}

We are familiar with the ADM charges of asymptotically flat spacetimes, which take the form of boundary intergrals computed at spatial infinity.  For `transverse asymptotically flat' spacetimes, in which the metric falls off to Minkowski spacetime only in directions perpendicular to a brane-like world volume, there are ADM spatial tensions in the brane directions \cite{Traschen:2001pb} as well as the ADM mass.  In the first law, the spatial tensions give the contribution of variations in the compactification lengths to the variation in the mass, as in (\ref{planarsmarr}).
The ADM mass and tensions of asymptotically planar AdS spacetimes were computed in \cite{El-Menoufi:2013pza}.  The purpose of this section is to carry out the corresponding computation for asymptotically planar dilaton-AdS spacetimes such as the Kasner-dilaton-AdS spacetimes which will be presented in section (\ref{dadsbhsection}) below.

With the falloff rates of these KDAdS  solutions  in mind, we take the approximate form of asymptotically-dilaton-AdS spacetimes at large radius to be
\begin{align}\label{asymptoticform}
ds^2 &\simeq \left({l^2\over r^2}\right)^{1-\sigma} (1+{c_r\over r^{n-\sigma}})dr^2 +{r^2\over l^2}\sum_{\alpha=0}^{n-1}(\epsilon_\alpha +{c_\alpha\over r^{n-\sigma}})(dx^{\alpha})^2\\
e^{a\phi}& \simeq \left({r^2\over l^2}\right)^{\sigma}(1+{ac_\phi\over r^{n-\sigma}})
\end{align}
where $\epsilon_0=-1$ and $\epsilon_k=+1$ for $k=1,\dots,n-1$, 
and $c_r$, $c_\alpha$, and $c_\phi$ are a set of dimensionful falloff coefficients that characterize the asymptotic form of the metric.
We show in  Appendix \ref{falloffappendix} that for solutions to the equations of motion for the action (\ref{action}) the falloff coefficients 
$(c_r,c_\alpha,c_\phi)$ for asymptotically planar dilaton-AdS spacetimes satisfy the relation
\begin{equation}\label{relation2}
c_r+\sum_{\alpha=0}^{n-1}\epsilon_\alpha c_\alpha -2ac_\phi= 0
\end{equation}
This generalizes a result for asymptotically planar AdS spacetimes found in \cite{El-Menoufi:2013pza}.  Note that the asymptotic form of the metric (\ref{asymptoticform}) is left invariant by the coordinate transformation
\begin{equation}\label{gauge}
r = r^\prime +{\lambda\over r^{\prime\, n-\sigma-1}}
\end{equation}
under which the falloff coefficients transform according to
\begin{equation}
c_r^\prime = c_r-2(n-2\sigma)\lambda,\qquad c_\alpha^\prime = c_\alpha +2\epsilon_\alpha\lambda,\qquad
c_\phi^\prime = c_\phi+{2\sigma\over a}\lambda
\end{equation}
which leaves the condition (\ref{relation2}) invariant.
We will want to check that the expressions for the ADM mass and spatial tensions are invariant under these gauge transformations.  

In order for the mass and spatial tensions to be finite, we take the spatial directions to be compact, with the identifications
\begin{equation}
x^k\equiv x^k+L_k
\end{equation}
and also define the volume $v=\prod_{k=1}^{n-1}L_k$.  As noted above, the compactification lengths $L_k$ are additional dimensionful quantities that are part of specifying a particular solution, and will play a role in the first law and Smarr formula for dilaton-AdS black holes. 

We compute the ADM mass and spatial tensions \cite{Traschen:2001pb} for asymptotically dilaton-AdS spacetimes using the Hamiltonian formalism.  The key new feature is the role played by the scalar field, which will yield a novel contribution to the mass and tensions.
The construction of an ADM charge takes place on a hypersurface $\Sigma$ with unit normal $n^a$, which will be taken alternatively to be either timelike or spacelike respectively  in computing the mass and spatial tensions.  The spacetime metric can then be decomposed as
\begin{equation}
g_{ab}=\epsilon n_an_b +s_{ab}
\end{equation}
where $\epsilon=n_an^a$ and $n^as_{ab}=0$, so that $s_{ab}$ is the metric on the hypersurface $\Sigma$.  The canonical momentum conjugate to $s_{ab}$ on $\Sigma$ is given by
\begin{equation}
\pi^{ab}=\epsilon \sqrt{|s|}(Ks^{ab}-K^{ab})
\end{equation}
where $K_{ab} = s_a{}^c\nabla_c n_b $ is the extrinsic curvature, while the momentum conjugate to the dilaton field $\phi$ is given by
\begin{equation}
p_{\phi}=-\epsilon \sqrt{|s|} n^a\nabla_a \phi.
\end{equation}  
and the Hamiltonian density for the scalar field is
\begin{equation}\label{scalarham}
{\cal H}_\phi = -\epsilon {F\over 2 \sqrt{|s|}} p_{\phi}^2 +p_{\phi}\beta^c D_c \phi 
+\sqrt{|s|}F\left({1\over 2} s^{ab} D_a \phi D_b \phi +{\cal V}(\phi) \right), \quad {\cal V}(\phi)=2\Lambda e^{-a\phi}
\end{equation}

We assume that the data on $\Sigma$ is asymptotic, at large radius, to that of the dilaton-AdS spacetime, which we denote by $(\bar s_{ab},\bar\pi^{ab},\bar{\phi},\bar{p}_{\phi})$.  The perturbations to the metric and the dilaton field and their corresponding momenta on $\Sigma$ are then given by
\begin{equation}
h_{ab} = s_{ab} -\bar s_{ab},\qquad p^{ab} = \pi^{ab}-\bar\pi^{ab},\qquad \delta\phi = \phi - \bar{\phi}, \qquad \delta p_{\phi} = p_{\phi} - \bar{p}_{\phi} 
\end{equation}

In Einstein-Maxwell theory there is a conserved electric charge, as well as the geometrical ADM charges. The electric charge arises due to the Gauss' law constraint, which enters the Maxwell Hamiltonian multiplied by the Lagrange multiplier $A_t$.
However the scalar field Hamiltonian (\ref{scalarham}) does not contain a constraint, and there is no separately
 conserved scalar charge. Instead, the ADM charges get a contribution from the scalar field, as follows.
 
To define the ADM charges one only needs a Killing field in the asymptotic region.
Let $\xi^a=F n^a +\beta^a$ be a Killing vector of the asymptotic dilaton-AdS spacetime, with $F$ and $\beta^a$ its projections normal and parallel to $\Sigma$. 
 The ADM charge associated with the asymptotic Killing vector $\xi^a$ evaluated at infinity on the slice $\Sigma$ is then given by
\begin{equation}\label{admcharge}
Q= - {1\over 16\pi}\int_{\partial\Sigma_\infty} da_cB^c= - {1\over 16\pi}\int_{\partial\Sigma_\infty} da_c(B^c_G+B^c_M)
\end{equation}
where the gravitational boundary vector $B^a_G$ is given by
\begin{equation}\label{bgrav}
B^a_G = F(D^a h-D_bh^{ab})-hD^aF +h^{ab}D_bF
+{1\over \sqrt{|s|}}\beta^b(\bar\pi^{cd}h_{cd}s^a{}_b-\bar\pi^{ac}h_{bc}-p^a{}_b)
\end{equation}
while the matter boundary vector $B^a_M$ is given by
\begin{equation}\label{bmatter}
B^a_M=(FD^a\phi+{1\over \sqrt{|s|}}\beta^ap_{\phi})\delta \phi
\end{equation}
and $D_a$ is the covariant derivative operator associated with the metric $\bar s_{ab}$ on $\Sigma$.

In order to compute the ADM mass, one takes $\Sigma$ to be a constant time slice and $\xi^a$ to be the time translation Killing vector of the asymptotic dilaton-AdS spacetime.  The key ingredients in the calculation of the boundary vector $B^a$ are  the projections of $\xi^a$ normal to and along $\Sigma$, which are given in this case by $F=r/l$ and $\beta^a=0$, and the components of the metric and dilaton perturbations on $\Sigma$, which has the non-vanishing components 
\begin{equation}
h_{rr} = \left({l^2\over r^2}\right)^{1-\sigma}{c_r\over r^{n-\sigma}},\qquad h_{kk} = \left({r^2\over l^2}\right){c_k\over r^{n-\sigma}}, \qquad
\delta\phi = {c_\phi\over r^{n-\sigma}}
\end{equation}
for $k=1,\dots,n-1$. 
Computations using these ingredients yield the expression for the ADM mass for asymptotic planar dilaton-AdS spacetimes, {\it i.e.}  the ADM charge $Q$ computed from the formula (\ref{admcharge}) with this choice for $\Sigma$ and $\xi^a$,
\begin{equation}\label{admmass}
M={v (n-1)\over 16\pi l^{n+1-\sigma}}\left( c_r + { (n-\sigma)\over (n-1) } \sum_{k=1}^{n-1}c_k-  a c_\phi\right)
\end{equation}
One can check that this expression is indeed invariant under the gauge transformation (\ref{gauge}) that preserves the asymptotically dilaton-AdS boundary conditions at infinity. 
One sees that the ADM mass is not determined solely in terms of the behavior of the metric at infinity, but depends on the decay of the scalar field towards its vacuum value as well. This contribution to $M$ vanishes if $a=0$,
and so is a consequence of the nontrivial behavior of the scalar field in the vacuum dilaton-AdS spacetimes (\ref{dilatonAdS}).  We expect that a similar effect happens with asymptotically Lifshitz boundary conditions \cite{Kachru:2008yh}, where matter fields similarly modify the asymptotic form of the vacuum spacetime\footnote{A scalar field contribution to the ADM mass also appears in reference \cite{Henneaux:2006hk}}.

In order to compute the ADM spatial tension in the $x^k$ direction, one takes a slice $\Sigma$ of constant $x^k$ and takes $\xi^a$ to be the spatial translation Killing vector of the asymptotic dilaton-AdS spacetime in the $x^k$ direction.  One again finds that the asymptotic Killing vector has a decomposition with respect to $\Sigma$ given by $F=r/l$ and $\beta^a=0$, while the metric and dilaton perturbations on $\Sigma$ now have the nonvanishing components
\begin{equation}
h_{rr} = \left({l^2\over r^2}\right)^{1-\sigma}{c_r\over r^{n-\sigma}},\qquad h_{\alpha\alpha} = \left({r^2\over l^2}\right){c_\alpha\over r^{n-\sigma}}, \qquad \delta\phi = {c_\phi\over r^{n-\sigma}}.
\end{equation}
for $\alpha=0,\dots,n-1$ with $\alpha\neq k$.  Again, after computations based on these ingredients one arrives at the expression for the ADM spatial tension ${\cal T}_k$  for asymptotic planar dilaton-AdS spacetimes\footnote{Strictly speaking this is the tension in the $x^k$ direction per unit time, because an integration over the time direction, which has not been periodically identified, has been suppressed.}
\begin{equation}\label{admtension}
{\cal T}_k={v (n-1)\over 16\pi l^{n+1-\sigma}L_k}\left( c_r + { (n-\sigma)\over (n-1) } 
\sum_{\alpha\neq k}\epsilon_\alpha c_\alpha-a c_\phi\right)
\end{equation}
%
One can check that this expression for the tension is also invariant under the gauge transformation (\ref{gauge}), and we see that it also receives a contribution from the falloff coefficient of the scalar field.
The expressions for the ADM mass and tensions are quite similar, with the ADM mass being essentially the tension in the time direction.

It is interesting to look at the `trace' of the ADM charges, which after using (\ref{relation2}) is given by
\begin{equation}
M +\sum_k {\cal T}_k L_k = a{(n-1)v\over 16\pi l^{n+1-\sigma}}\bigg\{n c_\phi-
 {a\over 2} (n-1) \sum_\alpha \epsilon_\alpha c_\alpha\bigg\}
\end{equation}
The right hand side vanishes for $a=0$, which gives the asymptotically AdS case, corresponding to tracelessness of the boundary CFT stress tensor.  From the bulk perspective this can be shown to follow from the existence of the scaling Killing vector (\ref{hkv}) in the AdS case \cite{El-Menoufi:2013tca}.  The failure of the trace of the charges to vanish for $a\neq 0$ then arises from hyperscaling violation.

\section{Dilaton-AdS black holes}\label{dadsbhsection}


In this section we present a class of plane symmetric solutions to (\ref{action}), which we call Kasner-dilaton-AdS (KDAdS) spacetimes which will serve as examples of asymptotically dilaton-AdS spacetimes and include a family of regular black hole solutions.  These are generalizations of Kasner-AdS spacetimes \cite{Linet:1986sr,Tian:1986zz,Sarioglu:2009vq,Ren:2016xhb}, which are solutions to (\ref{action}) with $a=0$ and vanishing scalar field. It will be useful to first recall these Kasner-AdS spacetimes in order to understand the properties of their dilatonic generalizations.  The Kasner-AdS spacetimes are given by
\begin{align}\label{kasner-AdS}
ds^2 & = {l^2\over r^2}{1 \over F(r)}dr^2+{r^2\over l^2}\sum_{\alpha=0}^{n-1}\epsilon_\alpha F^{p_\alpha}(r)(dx^\alpha)^2,\qquad 
F(r) =1-{r_0^n\over r^n}
\end{align}
%
where 
$r_0$ is a constant.
The exponents in (\ref{kasner-AdS}) satisfy the linear and quadratic Kasner constraints
\begin{equation}\label{kasner}
\sum_{\alpha=0}^{n-1} p_\alpha = \sum_{\alpha=0}^{n-1}  p_\alpha^2=1\end{equation}
Kasner-AdS spacetimes are symmetric under time translation, as well as translations in the spatial directons $x^k$.  However, the anisotropy introduced by the exponents $p_\alpha$ will generically break all boost and rotational symmetries in the $n$-dimensional Lorentzian plane spanned by the coordinates $(t, x^k)$.
Kasner-AdS spacetimes are asymptotic to AdS spacetime at large radius, while sufficiently close to $r=r_0$ they reduce to the vacuum Levi-Civita spacetimes\footnote{Lvei-Civita spacetimes are analytic continuations of vacuum Kasner cosmologies.}.  This can be seen by transforming to a new radial coordinate $y$ defined by
\begin{equation}
{r^n\over r_0^n}=1+{y^2\over y^2_0},\,\qquad y_0={2l\over n}
\end{equation}
and considering the regime $y\ll y_0$.  In this limit, the Kasner-AdS metric reduces to
\begin{equation}\label{lc-metric}
ds^2\simeq dy^2 + \sum_{\alpha=0}^{n-1}\epsilon_\alpha  \left({y\over y_0}\right)^{2p_\alpha} (dx^\alpha)^2
\end{equation}
%
This is the higher dimensional generalization of the cylindrically symmetric Levi-Civita metrics, which are solutions to the vacuum Einstein equations in four dimensions.  Except in certain special cases, the Levi-Civita spacetimes (\ref{lc-metric}) have curvature singularities at $y=0$.  This can be seen from the Kretschmann scalar, which is given by
\begin{equation}\label{kretschmann}
R_{abcd}R^{abcd} = {4\over y^4}\left ( \sum_\alpha p_\alpha^2(1-p_\alpha)^2 +\sum_{\alpha<\beta}p_\alpha^2 p_\beta^2\right)
\end{equation}
and given the Kasner constraints (\ref{kasner}), is clearly singular at $y=0$ unless one of the $p_\alpha$ equals $1$ with the rest vanishing, in which case $R_{abcd}R^{abcd}=0$.
If $p_0=1$ and $p_k=0$ for all $k$, the metric (\ref{lc-metric})  reduces to Minkowski spacetime in Rindler coordinates.   Taking a single spatial exponent {\it e.g.} $p_1=1$ with the remaining exponents vanishing  gives Minkowski spacetime in cylindrical coordinates.

Returning to the full Kasner-AdS metric (\ref{kasner-AdS}), choosing $p_0=1$ with $p_k=0$ for all $k$ gives the planar AdS black hole, which has a regular horizon at $r=r_0$
\begin{equation}\label{adsbh}
ds^2  ={l^2\over r^2} {dr^2\over F(r)} +{r^2\over l^2}\left(-F(r)dt^2+\sum_{k=1}^{n-1}(dx^k)^2\right) ,\qquad F(r) = 1-{r_0^{n-\sigma}\over r^{n-\sigma}}
\end{equation}
Regularity of the horizon follows from the regularity of the near horizon Levi-Civita limit (\ref{lc-metric}) for these values of the Kasner exponents.  Alternatively, taking a single spatial exponent {\it e.g.} $p_1=1$ with the remaining exponents vanishing gives an AdS solition \cite{Horowitz:1998ha}.   If the $x_1$ direction is periodically identified with the correct period, then the $x_1$ direction will close off smoothly at $r=r_0$.  So long as $p_0>0$, the surface $r=r_0$ in a Kasner-AdS spacetime will be a Killing horizon.  However, unless $p_0= 1$ this horizon will be singular as demonstrated by the near horizon limit (\ref{lc-metric}).  It has been argued in \cite{Ren:2016xhb} that more general Kasner-AdS spacetimes can nonetheless be physically relevant in the context of gauge-gravity duality.

We now want to consider a similar family of spacetimes that are solutions of (\ref{action}) and asymptotic at infinity to the dilaton-AdS spacetimes (\ref{dilatonAdS}).  We find that these `Kasner-dilaton-AdS spacetimes' are given by\footnote{It is shown in Appendix \ref{dimredappendix} that the Kasner-dilaton-AdS solutions to (\ref{action}) can be obtained by a generalized dimensional reduction in the sense of reference \cite{Gouteraux:2011qh} from generalized higher dimensional Kasner-AdS spacetimes.  The dilaton-AdS spacetimes (\ref{dilatonAdS}) are simply obtained by reduction from generalized higher dimensional AdS spacetime.}
\begin{align}\label{kdads}
ds^2& = {l^2\over r^2}{e^{a\phi}\over F}dr^2 
+{r^2\over l^2}\sum_{\alpha=0}^{n-1}\epsilon_\alpha F^{\hat p_\alpha}(dx^\alpha)^2,\qquad 
e^{a\phi}= \left({r\over l}\right)^{2\sigma} F^{{a\hat p_\phi/2}}\\  \nonumber
F(r)&=1-\left({r_0\over r}\right)^{n-\sigma}
\end{align}
where the `un-hatted' exponents $(p_\alpha,p_\phi)$ satisfy Kasner constraints augmented by an additional scalar exponent
\begin{equation}\label{modified}
\sum_{\alpha=0}^{n-1} p_\alpha = \sum_{\alpha=0}^{n-1}  p_\alpha^2+{1\over 2}p_\phi^2=1\end{equation}
and are related to the `hatted' exponents appearing in the metric and scalar field configurations (\ref{kdads}) by
\begin{equation}
\hat p_\alpha = {p_\alpha\over 1-a p_\phi/2},\qquad 
\hat p_\phi = {p_\phi\over 1-a p_\phi/2}
\end{equation}
The Kasner-dilaton-AdS spacetimes are asymptotic at large radius to the dilaton-AdS spacetime (\ref{dilatonAdS}).  For $a=0$, they give an extension of the Kasner-AdS including a massless scalar field that
 goes to zero at infinity, characterized by the additional Kasner exponent $p_\phi$.  

Massless scalar generalizations of the Levi-Civita spacetimes emerge in an expansion of the solutions (\ref{kdads}) around $r=r_0$.  Transforming to the new radial coordinate $y$ defined by 
\begin{equation}
\left({r\over r_0}\right)^{n-\sigma}=1+\left ({y^2\over y^2_0}\right)^{2(1-a p_\phi/2)},
\,\qquad y_0={2l(1-a p_\phi/2)\over n-\sigma}\left({r_0\over l}\right)^\sigma
\end{equation}
and restricting to the regime $y/y_0\ll 1$ one finds that
\begin{equation}\label{lc-metrictwo}
ds^2\simeq dy^2 + \sum_{\alpha=0}^{n-1}\epsilon_\alpha  \left({y\over y_0}\right)^{2p_\alpha} (dx^\alpha)^2,\qquad  
e^\phi = e^{\phi_0}\left({y\over y_0}\right)^{2p_\phi}
\end{equation}
with $e^{\phi_0} = (r_0/l)^{(n-1)a}$.  Since this metric is again of the Levi-Civita form, albeit with the modified constraints (\ref{modified}), the Kretschmann scalar is again given by (\ref{kretschmann}) and the analysis of curvature singularities at $y=0$ is very similar.  One finds that nonsingular geometries are possible only if the scalar exponent $p_\phi$ vanishes, in which case we again have the possible flat spacetime cases discussed above, in which one of the $p_\alpha$ equals $1$, with the remaining exponents vanishing.
It follows that the only cases for which the surface $r=r_0$ is regular in the full Kasner-dilaton-AdS metric (\ref{kdads}) are for the generalizations of the planar AdS black hole and AdS soliton with the scalar field. Taking $p_0=1$ and $p_k=0$ for all $k=1,\dots,n-1,\phi$ gives the dilaton-AdS black hole \cite{Charmousis:2010zz}
\begin{align}\label{dadsbh}
ds^2 & = \left( {l^2\over r^2}\right)^{1-\sigma} {dr^2\over F(r)} +{r^2\over l^2}\left(-F(r)dt^2+\sum_{k=1}^{n-1}(dx^k)^2\right) \\
\nonumber e^{a\phi} &= \left( {l^2\over r^2}\right)^{\sigma},\qquad F(r) = 1-{r_0^{n-\sigma}\over r^{n-\sigma}}
\end{align}
while the dilaton-AdS soliton is obtained by swapping the time direction with one of the planar spatial directions.

The Kasner-DAdS spacetimes (\ref{kdads}) have falloff coefficients
\begin{equation}\label{kdadsfalloff}
c_r = (1-{a\hat p_\phi\over 2})r_0^{n-\sigma},\quad
c_\alpha= -\epsilon_\alpha\hat p_\alpha r_0^{n-\sigma},\quad
c_\phi = -{\hat p_\phi\over 2}r_0^{n-\sigma}
\end{equation}
from which we determine that the ADM mass and spatial tensions for these spacetimes are given by
\begin{align}\label{kdads_mass}
M&= {v r_0^{n-\sigma}\over 16\pi l^{n+1-\sigma}}\left((n-\sigma)(\hat p_0-{a\over 2}\hat p_\phi)-(1-\sigma)\right)\\
{\cal T}_k &= {v r_0^{n-\sigma}\over 16\pi l^{n+1-\sigma}L_k}\left((n-\sigma)(\hat p_k-{a\over 2}\hat p_\phi)-(1-\sigma)\right)
\end{align}
Looking at the special case of the dilaton-AdS black hole (\ref{dadsbh}), which has $\hat p_0=1$ and $\hat p_k=\hat p_\phi=0$, we find that the mass and spatial tensions are given by
\begin{equation}
M = {(n-1)vr_0^{n-\sigma}\over 16\pi l^{n+1-\sigma}},\qquad  {\cal T}_k = -{(1-\sigma)vr_0^{n-\sigma}\over 16\pi l^{n+1-\sigma}L_k}
\end{equation}
which reduces in the limit $a=0$ to the correct result for the mass and tensions of the planar AdS black hole given in \cite{El-Menoufi:2013pza}.

\section{Smarr formula for dilaton-AdS black holes}\label{smarrsection}

In this section we derive the Smarr formula (\ref{planarsmarr}) using Komar integrals\footnote{The $a=0$ limit of this derivation gives a more straightforward construction of the Smarr formula for planar AdS black holes established in \cite{El-Menoufi:2013pza}.}, which provide a Gauss's law type statement for black hole spacetimes with an exact Killing symmetry, that relates a boundary integral at spatial infinity to one evaluated at the black hole horizon. We consider spacetimes which
are aysmptotic to the dilaton AdS metric (\ref{dilatonAdS}) and have regular black hole horizons. Further, we will show that the result 
holds on spacetimes that do not have regular black hole horizons, but whose singular behavior is described by the 
KDAdS metrics (\ref{lc-metrictwo}). The only proviso in this case is that depending on the choice of the Kasner exponents, either
 $\kappa$ or $A$ is infinite. However, their product is finite, resulting in a sensible contribution from the interior 
 singular surface.

The starting point for the construction is Einstein's equation for the theory (\ref{action}), which we can write as
\begin{equation}
R_{bc}={1\over 2}(\nabla_b\phi)\nabla_c\phi + {2\Lambda\over D-2}g_{bc}e^{-a\phi}
\end{equation}
It follows that for a solution to these equations with a Killing vector $\xi^b$, the Killing vector satisfies
\begin{equation}\label{identity2}
\nabla_b\nabla^b\xi^c =- {2\Lambda\over D-2}e^{-a\phi}\xi^c
\end{equation}
where we have assumed that the Killing vector is also a symmetry of the scalar field, so that $\xi^a\nabla_a\phi=0$.
The right hand side of (\ref{identity2}) can be rewritten in terms of a generalized Killing potential $\omega^{ab}$ defined by\footnote{See \cite{Kastor:2009wy} for a discussion of Killing potentials.}
\begin{equation}\label{genkilling}
 \nabla_b(e^{-a\phi}\omega^{bc})=e^{-a\phi}\xi^c
\end{equation}
The integrability condition for the existence of such a generalized Killing potential
\begin{equation}
\nabla_b(e^{-a\phi}\xi^b) = e^{-a\phi}\nabla_b\xi^b -a e^{-a\phi}\xi^b\nabla_b\phi=0
\end{equation}
is satisfied by virtue of Killing's equation and the assumption that $\xi^a$ is a symmetry of the scalar field.  Equation (\ref{identity2}) can now be written as
\begin{equation}
\nabla_b(\nabla^b\xi^c+{2\Lambda\over D-2}e^{-a\phi}\omega^{bc}) =0
\end{equation}
Let $\Sigma$ be a codimension-1 slice of the spacetime with unit normal $n_a$ stretching between the black hole horizon and infinity.  Gauss's law then relates boundary integrals at infinity and the black hole horizon $I_\infty= I_H$ with
\begin{equation}
I_k =\int_{\partial\Sigma_k} da_{bc}(\nabla^b\xi^c +{2\Lambda\over D-2}e^{-a\phi}\omega^{bc})
\end{equation}
The volume form is given by $da_{bc} = da \rho_{[b}n_{c]}$ where $da$ is the intrinsic volume element on $\Sigma$ and  
$\rho_b$ is the outward pointing unit normal to the boundary component $\partial\Sigma_k$.

Now consider the case $\xi=\partial/\partial t$ with $\Sigma$ taken to be contant time slice.  As usual in the Komar derivation of the Smarr formula, the Killing vector term in the boundary integral at the horizon is proportional to $\kappa A$, yielding
\begin{equation}
I_H = -2\kappa A +  {2\Lambda\over D-2}\int_{\partial\Sigma_H}da_{bc}\, e^{-a\phi}\omega^{bc}
\end{equation}
At infinity, we find that the contributions from both the Killing vector and generalized Killing potential terms are divergent, but cancel leaving a finite result.  To evaluate these terms, we use the asymptotic form of the metric (\ref{asymptoticform}), which gives for the relevant components and to the necessary order in $1/r$
\begin{equation}
\nabla^r\xi^t\simeq \left({l^2\over r^2}\right)^\sigma\left\{{r\over l^2}+{(n-\sigma)c_t-2c_r\over 2l^2 r^{n-\sigma-1}}\right\}
\end{equation}
For the generalized Killing potential, one finds that 
\begin{equation}
\omega^{rt} \simeq {r\over n-\sigma}+{\alpha\over r^{n-\sigma-1}}
\end{equation}
where $\alpha$ is an arbitrary (dimensionful) constant, which corresponds to adding a homogeneous solution to (\ref{genkilling}) to the generalized Killing potential.  Recalling that $\Lambda = -(n-\sigma)(n-1)/2l^2$, one sees that the divergent terms in $I_\infty$ will cancel.  Further, if we define the antisymmetric quantity $\omega^{rt}_{div}$ to have the non-zero components
\begin{equation}
\omega^{rt}_{div} = {r\over n-\sigma}-{(1+\sigma)c_r+(n-2\sigma-1)ac_\phi\over 2(n-\sigma)r^{n-1}}
\end{equation}
one finds, after making use of the relation (\ref{relation2}) between the falloff coefficients, the gauge invariant result
\begin{equation}
\int_{\partial\Sigma_\infty} da_{bc}(\nabla^b\xi^c+{2\Lambda\over D-2}e^{-a\phi}\omega_{div}^{bc}) = -16\pi M
\end{equation}
Defining the quantity $\Psi$ by
\begin{equation}\label{psidef2}
\Psi \equiv \int_{\partial\Sigma_H}da_{bc}\,e^{-a\phi}\omega^{bc}
- \int_{\partial\Sigma_\infty}da_{bc}\,e^{-a\phi}(\omega^{bc}-\omega_{div}^{bc})
\end{equation}
and collecting terms, the equality $I_\infty=I_H$ between boundary terms yields the Smarr relation
\begin{equation}\label{komarsmarr2}
(D-2) M = (D-2){\kappa A\over 8\pi} - 2 {\Psi\Lambda\over 8\pi}
\end{equation}
We can compare this result with the Smarr formula given in (\ref{planarsmarr}) and see that these two expressions differ in certain dimensional factors, in the presence of tension terms in (\ref{planarsmarr}) not present in (\ref{komarsmarr2}), and in the possible difference between the quantities $\Theta $ and $\Psi$.
The two formulas can be reconciled by adding and subtracting the quantity $M+\sum_k{\cal T}_kL_k$ to the right hand side of (\ref{komarsmarr2}), giving the relation
\begin{equation}\label{finalsmarr}
(D-3)M = (D-2){\kappa A\over 8\pi} +\sum_k {\cal T}_kL_k - 2 {\Theta\Lambda\over 8\pi}
\end{equation}
where the thermodynamic potential $\Theta$ is now related to the integrals (\ref{psidef2}) of the generalized Killing potential
$\Psi$ according to
\begin{equation}\label{reconcile}
\Theta = \Psi + {4\pi\over\Lambda}(M+\sum_k{\cal T}_kL_k)
\end{equation}
We will see that the same relation between $\Theta$ and $\Psi$ emerges from the derivation of the first law.  Note that in the AdS case, with $a=0$, the trace of the ADM charges vanishes, $M+\sum_k{\cal T}_kL_k=0$, and hence the quantities $\Theta$ and $\Psi$ are equal, as in \cite{El-Menoufi:2013pza}.
Equation (\ref{reconcile}) can now be used to compute the thermodynamic potential $\Theta$ for the dilaton-AdS black holes (\ref{dadsbh}) with the result
\begin{equation}\label{finaltheta}
\Theta = -{vr_0^{n-\sigma}\over 2(n-\sigma)l^{n-\sigma-1}}
\end{equation}
We discuss the thermodynamic meaning of $\Theta$ in Section (\ref{phenosection}).

Let us also consider the Smarr formula (\ref{finalsmarr}) for the broader class of singular KDAdS spacetimes (\ref{kdads}). The mass and tensions for the KSAdS spacetimes are well defined and were computed in (\ref{kdads_mass}).  One might 
expect that the interior boundary term and Killing potential contributions to the Smarr formula should diverge. However, it is 
straightforward to check, using the 
coordinates of (\ref{lc-metrictwo}), that the interior boundary term contribution from $\nabla_a \xi_b$ is finite, that the integral of
the particular solution for $\omega^{yt}$ vanishes, and that any possible contribution from a homogeneous solution for the generalized Killing potential is
finite. Hence, all terms in the Smarr relation (\ref{finalsmarr}) are well defined for the entire KDAdS family of solutions, with 
the subtlety that the term $\kappa A$ is defined only as a product.

\section{First law for dilaton-AdS black holes}\label{firstlawsection}

We derive the first law for dilaton-AdS black holes using
Hamiltonian perturbation theory formalism set out in  \cite{Traschen:2001pb,El-Menoufi:2013pza}, 
which we will set out a brief account of here. 
For this purpose we consider perturbing from a static, spatial translation invariant dilaton-AdS black hole solution to a nearby solution, which is not required to be static or spatial translation invariant.   We also include the possibility of varying the compactification lengths $L_k$, as well as the strength $\Lambda$ of the scalar potential (or equivalently the dilaton-AdS length scale $l$).  We make use of the Hamiltonian variables and $n+1$ decomposition introduced in Section \ref{admsection}
in order to define the ADM charges. Applying the techniques of \cite{Traschen:2001pb,El-Menoufi:2013pza} to 
the Einstein plus scalar field system one can show that solutions $( h_{ab} , p^{ab} , \delta \phi, \delta p_\phi )$
to the linearized field equations about a spacetime $g_{ab}^{(0)}$ with Killing field $\xi^a =Fn^a + \beta^a$ obey
a Gauss Law type relation,
\begin{equation}\label{gauss}
D_a B^a =0
\end{equation}
where the vector $B^a = B_G^a + B_M^a$ is given in equations (\ref{bgrav}) and (\ref{bmatter}). 

Taking $\xi^a$ to be the time translation Killing field,
the perturbations to the metric and scalar field near infinity will then be given by\footnote{See reference \cite{Kastor:2006ti} for a discussion of the $\delta L_k$ terms in the metric perturbation.}
\begin{align}
h_{rr}&\simeq \left({l^2\over r^2}\right)^{1-\sigma}{\delta c_r\over r^{n-\sigma}}
+\left({l^2\over r^2}\right)^{1-\sigma}(1+{c_r\over r^{n-\sigma}}){2(1-\sigma)\delta l\over l}                    \\
h_{kk}&\simeq {r^2\over l^2}\,{\delta c_k\over r^{n-\sigma}} + {r^2\over l^2}(1+{c_k\over r^{n-\sigma}})
\left( {2\delta L_k\over L_k} - {2\delta l\over l}\right)\\
\delta\phi&\simeq {\delta c_\phi\over r^{n-\sigma}} - (n-1)a{\delta l\over l}
\end{align}
The Gauss law (\ref{gauss}) can be integrated over the slice $\Sigma$, giving a relation between a boundary term at infinity to one at the horizon, $J_\infty=J_H$, where
\begin{equation}
J_k = \int_{\partial\Sigma_k} da_c \left(B^c +2(n_an^a)e^{-a\phi}\omega^{cd}n_d\delta\Lambda\right)
\end{equation}
where the generalized Killing potential $\omega^{ab}$ was defined in (\ref{genkilling}).
The volume element is given by $da_c = da\rho_c$, where $da$ is the intrinsic volume element on the boundary and $\rho_c$ is the outward pointing unit normal to the boundary within $\Sigma$.

The horizon boundary integral of $B^a$ is proportional to the change in horizon area under the perturbation, so that the boundary term at the horizon may be written as
\begin{equation}
J_H = -2\kappa\delta A -{2(n-1)(n-\sigma)\over l^2}{\delta l\over l} \int_{\partial\Sigma_k} da_c e^{-a\phi}\omega^{cd}n_d
\end{equation}
The radial component of the boundary vector $B^a$ near infinity is found to be given by
\begin{align} \nonumber
B^r  \simeq & {n-1\over l}\left(r^2\over l^2\right)^{1-\sigma} \Bigg  \{ -2 {\delta l\over l}     
 +{1\over r^{n-\sigma}}\bigg [ -\delta c_r +{n-\sigma\over n-1 } \sum_k\delta c_k -a\delta c_\phi      
-{n-\sigma \over n-1 } \sum_k (c_t+c_k){\delta L_k\over L_k}  \\   
+ & \Big( 2c_r +(2-\sigma)\sum_k c_k + (n+1-\sigma)c_t + (n-\sigma)ac_\phi\Big){\delta l\over l}\bigg ] \Bigg \}  \label{firstboundary}
\end{align}
The first term in (\ref{firstboundary}) diverges, but this divergence is cancelled by a contribution from the Killing potential term near infinity, which is given by
\begin{equation}
2e^{-a\phi}\omega^{rt}_{div}n_t\delta\Lambda\simeq
\left({r^2\over l^2}\right)^{1-\sigma}{\delta l\over l^2}\bigg\{ -2(n-1) 
+{(n-1)\over r^{n-\sigma}}\Big[ c_t+(1+\sigma)c_r +(n-2\sigma+1)a c_\phi\Big ]\bigg\}
\end{equation}
These two terms then combine to give the finite boundary term at infinity
\begin{align}
J_\infty = & -16\pi \bigg( \delta M -\sum_k{\cal T}_k\delta L_k +(M+  \sum_k{\cal T}_k L_k ){\delta l\over l }   \bigg)\\
& \nonumber -{2(n-1)(n-\sigma)\over l^2}{\delta l\over l}\int_\infty da_c e^{-a\phi}( \omega^{cd}- \omega^{cd}_{div})n_d
\end{align}
Further combining this result with $J_H$ then gives the first law
\begin{equation}\label{finalfirstlaw}
\delta M = {\kappa\delta A\over 8\pi}+\sum_k{\cal T}_k \delta L_k +{\Theta\delta\Lambda\over 8\pi}
\end{equation}
where in agreement with the results of the last section, the thermodynamic potential $\Theta$ is given by
\begin{equation}
\Theta =  \Psi - {8\pi l^2\over (n-1)(n-\sigma)}(M+  \sum_k{\cal T}_k L_k )
\end{equation}

With a pure cosmological constant, {\it i.e.} with $a=0$, it is common to consider the quantity $P=-\Lambda/8\pi$ to be a pressure, and one can then interpret
its thermodynamic conjugate $V=-\Theta $ in the first law (\ref{firstlaw}) to be a thermodynamic volume.  This leads in turn to the interpretation of $M$ as the enthalpy of the black hole spacetime \cite{Kastor:2009wy}.   We have shown that the first law for dilaton-AdS black holes (\ref{finalfirstlaw}) has the same form for $a\neq 0$ with the overall strength 
of the scalar potential, $\Lambda$, playing the role of the cosmological constant.  It is thus also natural to consider $M$ as measuring the enthalpy of the dilaton-AdS black hole spacetime, justifying the title of this paper.

However, it is worth noting that with $a\neq 0$, the interpretation of $\Lambda$ as pressure is somewhat less direct.
In this case, viewing the scalar field stress-energy as that of a perfect fluid, the pressure in the radial direction is given by $P_r ={1\over 2} g^{rr} (\partial_r\phi )^2 \  -{\cal V}(\phi)$, while the 
transverse pressures and energy density are given by $P_k =-\rho = -{1\over 2} g^{rr} (\partial_r\phi)^2 \  -{\cal V}(\phi)$.   There is no single `pressure', even at a given spacetime point.  However, in a
``slow roll" limit such that the scalar field evolves gradually in the radial direction and the potential ${\cal V}(\phi)$ dominates
over the gradient term in the stress energy, then ${\cal V}(\phi)$ is a slowly (radially) evolving effective cosmological constant.   From the dilaton-AdS metric (\ref{dilatonAdS}), we see that the slow roll condition holds when $a^2 \ll 1$.
Hence in this limit, a variation in $\Lambda$ leads proportionally to a variation in the actual spacetime pressure at each point.

\section{Explicit thermodynamic check}\label{phenosection}


In this section, we check that the dilaton-AdS black hole solution (\ref{dadsbh}) indeed satisfies the first law (\ref{finalfirstlaw}) and Smarr formula (\ref{finalsmarr}).
We have seen that the mass and tensions for the planar DAdS black hole are given by
\begin{equation}\label{dads_bh_mass}
M= {(n-1)vr_0^{n-\sigma}\over 16\pi l^{n+1-\sigma}},\qquad
{\cal T}_k = -{(1-\sigma)vr_0^{n-\sigma}\over 16\pi l^{n+1-\sigma}L_k}
\end{equation}

The entropy of the planar DAdS black hole, equal to a quarter of the area of the event horizon, and the temperature, equal to the surface gravity over $2\pi$, are found to be
\begin{equation}\label{sandt}
S = {vr_0^{n-1}\over 4 l^{n-1}},\qquad T= {(n-\sigma)r_0^{1-\sigma}\over 4\pi l^{2-\sigma}}
\end{equation}
%
In order to explore the thermodynamics of the planar DAdS black hole in a phenomenological way, we can write the mass as a function of a set of independent variables given by the entropy $S$, compactification lengths $L_k$, and DAdS length scale $l$.  This works out to be
\begin{equation}
M(S,L_k,l) = {n-1\over 16\pi l}(4S)^{{n-\sigma\over n-1}}v^{-{1-\sigma\over n-1}}
\end{equation}
where $v=\prod_{k=1}^{n-1}L_k$ is the volume of the compact directions.  We expect that the mass should satisfy the first law
\begin{equation}
\delta M = T\delta S +\sum_{k=1}^{n-1}{\cal T}_k \delta L_k + {\Theta \over 8\pi}\delta \Lambda
\end{equation}
where $\Lambda = -{(n-1)(n-\sigma)\over 2l^2}$.  One finds that the temperature $T$ in (\ref{sandt}) and spatial tensions ${\cal T}_k$ in (\ref{dads_bh_mass}) are reproduced respectively by the partial derivatives 
$(\partial M/\partial S)_{L_k,\Lambda}$ and $(\partial M/\partial L_k)_{S,L_{j\neq k},\Lambda}$.
The thermodynamic conjugate of $\Lambda$ is found to be
\begin{equation}\label{thermov}
\Theta = 8\pi \left({\partial M\over \partial \Lambda}\right)_{S,L_k} = -{vr_0^{n-\sigma}\over 2(n-\sigma)l^{n-1-\sigma}}
\end{equation}
in agreement with the result (\ref{finaltheta}) found from our expression for the thermodynamic potential (\ref{reconcile}).
The thermodynamic volume is given by
\begin{equation}
V=-\Theta = {vr_0^{n-\sigma}\over 2(n-\sigma)l^{n-1-\sigma}}
\end{equation}
%
The thermodynamic volume is simply related in this case, as in some of the examples in \cite{Cvetic:2010jb},  to the integral of the scalar potential over the volume behind $r=r_0$ given by
\begin{equation}
W = \int_{r=0}^{r_0}drd^{n-1}x\sqrt{-g}\, {\cal V}(\phi) = v \int_{r=0}^{r_0}dr\left({r\over l}\right)^{n-1+\sigma}\left({r\over l}\right)^{-2\sigma} = 
{vr_0^{n-\sigma}\over (n-\sigma)l^{n-1-\sigma}}
\end{equation}
Hence the thermodynamic volume is proportional to this ``interior" average of the scalar field potential
\begin{equation}
V ={1\over 2}W
\end{equation}
In the pure cosmological constant case, $a=0$, the thermodynamic volume is just one-half the Euclidean volume ``behind" the black hole. This fits in nicely with understanding the cosmological constant contribution
to the Smarr formula as a $PV$-type contribution to the free energy, as in classical thermodynamics. 
The scalar field action in (\ref{action}) with an exponential potential  can be shown to arise via generalized dimensional reduction \cite{Gouteraux:2011qh} from pure Einstien gravity with a cosmological
constant in a higher (not necessarily integer) dimension, as shown in Appendix \ref{dimredappendix}.
Hence, in the $a\ne 0$ case $V$ can also be thought of as the interior volume of a generalized higher dimensional black hole

\section{Conclusions}\label{conclusions}

In this paper, we have derived expressions for the ADM mass and tensions of spacetimes that are asymptotic at infinity to the dilaton-AdS family of vacua (\ref{dilatonAdS}) that arise as solutions to the coupled Einstein-scalar field system (\ref{action}).  The system includes an exponential potential for the scalar field that generalizes the case of a pure cosmological constant.  As a consequence of the nontrivial behavior of the scalar field in the vacuum, we find that the mass and tensions depend on the falloff behavior near spatial infinity of the scalar field as well as the metric.  We went on to prove an extended first law (\ref{finalfirstlaw}) for asymptotically dilaton-AdS black holes  that includes variations with respect to the overall strength $\Lambda$ of the scalar potential.  We also proved, both using Komar integrals and via a scaling argument from the extended first law, a Smarr formula for these black holes (\ref{finalsmarr}) that includes the contributions from the scalar potential.  These results are very much in line with those obtained in \cite{Kastor:2009wy} in the case of a pure cosmological constant.  This demonstrates that the response of black holes to changes in their environment in this more complicated system still takes a similar form.

There are a number of interesting extensions of these results that we would like to pursue.  First, there is a deSitter analogue of the Einstein-scalar field system (\ref{action}) in which $\Lambda$ is taken to be positive.  Solutions with $\Lambda >0$, corresponding to those studied here, are cosmologies that can be obtained via analytic continuation from the dilaton-AdS vacua (\ref{dilatonAdS}) and the Kasner-DAdS solutions (\ref{kdads}).  In particular, the slow roll limit provides interesting inflationary models.
The computation of masses and tensions presented here analytically continue into cosmic hair, in the sense of \cite{Kastor:2016bnm}, for such slow-roll inflationary scenarios.
Second, matter fields with non-trivial behavior at infinity are present in Lifshitz vacua \cite{Kachru:2008yh} and in vacua that combine anisotropic Lifshitz scaling with hyperscaling violation \cite{Dong:2012se}.  The techniques used here should also be effective in exploring the extended thermodynamics \cite{Brenna:2015pqa,Ma:2017pap} of black holes in these systems \cite{Taylor:2008tg,Danielsson:2009gi,Mann:2009yx,Bertoldi:2009vn}.

%

\subsection*{Note Added}  After work on the present paper was largely complete, we became aware of overlapping work in \cite{pedraza}, which in addition deals with anisotropic Lifshitz scaling and different horizon topologies.

\subsection*{Acknowledgements}  The work of S.R. is supported by FONDECYT grant  1150907. 

\vskip 0.3in

\appendix

{\huge\bf  Appendix}

\section{Relation for falloff coefficients}\label{falloffappendix}

We demonstrate that solutions to the field equations for the action (\ref{action}) that satisfy the asymptotically dilaton-AdS falloff conditions (\ref{asymptoticform}) at large radius have falloff coefficients satisfying the relation (\ref{relation2}).  The derivation parallels that given in \cite{El-Menoufi:2013pza} for the asymptotically AdS case.  The field equations for (\ref{action}) are given by
\begin{align}
&R_{ab}-{1\over 2}(\nabla_a\phi)\nabla_b\phi - {2\Lambda\over D-2}g_{ab}e^{-a\phi}=0\\
&\nabla^2\phi+2a\Lambda e^{-a\phi}=0
\end{align}
and the trace of the gravitational field equation is given by
\begin{equation}
R-{1\over 2}(\nabla\phi)^2 -{2\Lambda D\over D-2}e^{-a\phi}=0
\end{equation}
Consider this trace equation linearized around the asymptotic dilation-AdS background (\ref{dilatonAdS}) in the large radius regime.  Let us write the metric and scalar field as
\begin{equation}
g_{ab}=\bar g_{ab} +\gamma_{ab},\qquad \phi = \bar\phi + \delta\phi
\end{equation}
where $(\bar g_{ab},\bar\phi)$ are the asymptotic background fields.  For spacetimes satisfying the asymptotically dilaton-AdS falloff conditions (\ref{asymptoticform}) in the large radius regime, we have
\begin{align}
\gamma_{rr}&=\left({l^2\over r^2}\right)^{1-\sigma}{c_r\over r^{n-\sigma}},\qquad
\gamma_{\alpha\alpha}=\left({r^2\over l^2}\right) {c_\alpha\over r^{n-\sigma}}\\
\delta\phi &={c_\phi\over r^{n-\sigma}}
\end{align}
The linearization of the traced field equation, after making use of the background field equations, can be reduced to
\begin{equation}
-\bar\nabla^2\gamma +\bar\nabla_a\bar\nabla_b\gamma^{ab}  -\bar\nabla_a(\delta\phi\bar\nabla^a\bar\phi)
 -{2\Lambda\over D-2}e^{-a\phi}(\gamma-2a\delta\phi)=0
\end{equation}
where $\bar\nabla_a$ is the background covariant derivative operator and $\gamma=\bar g^{ab}\gamma_{ab}$.  Calculation then shows that the first three terms each vanish identically and one is left with the equation
\begin{equation}
\gamma-2a\delta\phi=0
\end{equation}
which reduces to the relation (\ref{relation2}) between the falloff coefficients.

\section{Dilaton AdS spacetimes via generalized dimensional reduction}\label{dimredappendix}

The dilaton-AdS spacetimes (\ref{dilatonAdS}) and Kasner-DAdS spacetimes (\ref{kdads}) may be obtained via generalized dimensional reduction \cite{Gouteraux:2011qh} starting from higher (non-integer) dimensional AdS and Kasner-AdS spacetimes respectively. One starts with the action
\begin{equation}\label{genaction}
S_{N+1} = {1\over 16\pi G_{N+1}}\int d^{N+1}x \sqrt{-g_{N+1}}(R-2\Lambda)
\end{equation}
where $N$ will generally be non-integer.  Dimensionally reducing according to
\begin{equation}\label{genreduction}
ds^2_{N+1} = e^{-a\phi}ds^2_{n+1} +e^{(a_c^2-a^2)\phi/ a}dX_{N-n}^2
\end{equation}
where $dX_{N-n}^2$ is the Euclidean metric in $N-n$ dimensions and these directions have all been identified with period $L_c$, yields the dimensionally reduced action
\begin{equation}\label{redaction}
S_{n+1} = {1\over 16\pi G_{n+1}}\int d^{n+1}x\sqrt{-g}(R-{1\over 2}(\nabla\phi)^2 - 2\Lambda e^{-a\phi})
\end{equation}
with $G_{N+1}=L_c^{N-n}G_{n+1}$, which matches the action (\ref{action}) for Einstein gravity coupled to a scalar field with an exponential potential.   The coupling $a$ and the parameter $a_c$ appearing in the dimensional reduction are given in terms of $N$ and $n$ by
\begin{equation}
a^2 = {2(N-n)\over (n-1)(N-1)},\qquad a_c^2 = {2\over n-1}
\end{equation}
We can start with $(N+1)$-dimensional `generalized' AdS spacetime
\begin{align}
ds_{N+1}^2 &= {L^2\over R^2} dR^2 + {R^2\over L^2}\eta_{AB}dx^Adx^B, \qquad A,B = 0,1,\dots,N-1\\
\Lambda &= -{N(N-1)\over 2L^2}\nonumber
\end{align}
Dimensional reduction via (\ref{genreduction}) then yields
\begin{align}
ds_{n+1}^2 &= e^{a\phi}\,\left( {L^2\over R^2} dR^2 + {R^2\over L^2}\eta_{\alpha\beta}dx^\alpha dx^\beta\right),
\qquad \alpha,\beta = 0,1,\dots,n-1 \\
e^{a\phi} &= \left({R\over L}\right)^{2\gamma},
\qquad  \gamma = {N-n\over n-1}\nonumber
\end{align}
The form of the dilaton-AdS spacetimes (\ref{dilatonAdS}) is obtained by redefining the radial coordinate according to
\begin{equation}\label{redefR}
{R\over L} = \left({r\over l}\right)^{1-\sigma}
\end{equation}
where as in the main part of the paper $\sigma = (n-1)a^2/2$ and the DAdS length scale $l$ is given in (\ref{dlambda}).

In order to obtain the Kasner-dilaton-AdS (KDAdS) spacetime (\ref{kdads}), which includes the dilaton-AdS black hole (\ref{dadsbh}), we start with $(N+1)$-dimensional Kasner-AdS spacetime given by
\begin{align}
ds^2 & = {L^2\over R^2}{1 \over F(R)}dr^2+{R^2\over L^2}\sum_{\alpha=0}^{N-1}\epsilon_\alpha F^{P_\alpha}(R)(dx^\alpha)^2,\qquad 
F(R) =1-{R_0^N\over R^N}\\
\Lambda &= -{N(N-1)\over 2L^2}\nonumber
\end{align}
with Kasner exponents $P_i=P$ for $i=n,\ldots,N-1$. The constraints (\ref{kasner}) on the exponents then take the form
\begin{align}\label{Kasnercons}
\sum\limits_{\alpha=0}^{n-1}P_{\alpha}+(N-n)P=1=\sum\limits_{\alpha=0}^{n-1}P_{\alpha}^2+(N-n)P^2.
\end{align}
Dimensional reduction (\ref{genreduction}) then gives
\begin{align}
ds_{n+1}^2 &= e^{a\phi}\,\left( {L^2\over R^2F(R)} dR^2 + {R^2\over L^2}\sum_{\alpha=0}^{n-1}\epsilon_\alpha F^{P_\alpha}(R)(dx^\alpha)^2\right),
\qquad \alpha= 0,1,\dots,n-1 \\
e^{a\phi} &= \left({R\over L}\right)^{2\gamma}F^{\gamma P}(R),
\qquad  \gamma = {N-n\over n-1}\nonumber
\end{align}
As previously, redefining the radial coordinate according to (\ref{redefR}) brings the metric to the form
\begin{align}
ds_{n+1}^2 &={l^2\over r^2}{e^{a\phi} \over f}dr^2+{r^2\over l^2}\sum_{\alpha=0}^{n-1}\epsilon_\alpha f^{P_\alpha+\gamma P}(dx^\alpha)^2\\
f(r)&=1-\left(\dfrac{r_0}{r}\right)^{n-\sigma}\\
e^{a\phi} &= \left({r\over l}\right)^{2\sigma}f^{\gamma P}(r)
\end{align}
Finally, redefining the exponents on $f(r)$ as
\begin{align*}
\hat p_\phi={p_\phi\over 1-a p_\phi/2}={2\gamma P\over a} \quad \text{and} \quad \hat p_\alpha={p_\alpha\over 1-a p_\phi/2}=P_\alpha+\gamma P
\end{align*}
yields the metric (\ref{kdads}) while the constraints (\ref{Kasnercons}) transform into (\ref{modified}).

\end{document}